\documentstyle[aps,pra,amsmath,multicol]{revtex}
\input{epsf.tex}
\epsfverbosetrue
\begin{document}
\title{Modulational instability of spinor condensates}
\author{Nicholas P. Robins$^1$, Weiping Zhang$^{1,2}$, Elena A. Ostrovskaya$^1$, 
and Yuri S. Kivshar$^1$}
\address{$^1$Nonlinear Physics Group, Research School of Physical Sciences 
\& Engineering,\\ The Australian National
University, Canberra ACT 0200, Australia \\
$^2$Optical Sciences Center, University of Arizona, Tuscon, AZ 85721, USA}
\maketitle
\begin{abstract}
 We demonstrate, analytically and numerically, that the ferromagnetic 
phase
 of the spinor Bose-Einstein condenstate  may experience modulational instability of the 
ground 
state leading to a fragmentation of the spin domains.  Together with other 
nonlinear effects in the atomic optics of ultra-cold gases (such as 
coherent photoassociation and four-wave mixing) this effect provides one 
more analogy between coherent matter waves and light waves
 in nonlinear optics. 
\end{abstract}

\begin{multicols}{2}
Recent experimental studies of Bose-Einstein 
condensation (BEC) in an
optical trap \cite{exp} have opened a new direction in 
the research 
of ultra-cold atomic clouds associated with their {\em spin degree of 
 freedom},
 or {\em spinor BEC}.  An optical trap does not force atoms to align 
along the 
orientation of the strong confining magnetic field, as happens in the 
case 
of a magnetic trap, allowing the study of atoms confined in all 
hyperfine 
states. Several recent theoretical studies have predicted a variety 
of novel 
phenomena that may occur in the spinor BEC, such as the propagation of 
spin waves and the existence of topological states -{\em skyrmions}, vortex states 
without a core \cite{ho,law,pu,yip,japan,weiping,japan2}.
In the case of spin-1 bosons such as $^{23}Na$, $^{39}K$ and 
$^{87}Rb$,
 the dynamics of the spinor BEC is described by the three spin 
degrees 
of freedom ($m_F=1,0,-1$ of the $F=1$ atomic hyperfine state) which are 
coupled parametrically.  Depending on the parameters, such as
scattering
 lengths, in an optical trap the ground state of the spinor BEC can be
 either {\em ferromagnetic} or {\em antiferromagnetic} (''polar'').

  In this
 paper, we demonstrate that the parametric coupling between the 
spin degrees 
of freedom provides a physical mechanism for the {\em modulational 
instability}
 of the ferromagnetic ground state, for large enough densities of 
spinor BEC.
 In contrast, the antiferromagnetic state is always modulationally stable. 
 This effect is reminiscent of the quasiparticle instabilities in two-component homogeneous condensates that are known to occur only for certain ratios of the inter- and intra-component interaction strengths \cite{goldstein}.  It also suggests one more example of a deep analogy between
 the coherent matter waves and light waves in nonlinear optics, along
 with the already studied cases of four-wave mixing \cite{fwm} and 
atomic-molecular
 photoassociation \cite{ambec}.

{\em Model.} We consider an atomic spinor BEC in an optical 
trap in the 
magnetic-field-free case.  The Hamiltonian for the spinor BEC in the 
optical 
 trap has the form \cite{ho}:
\begin{equation}
 \label{Hamiltonian}
H=\int d{\bf r} \left( S_{m_F}+
\frac{c_0}{2}S_{m_{F,F'}}+
\frac{c_2}{2}S_{m^{a,a'}_F,m^{b,b'}_F}\right),
\end{equation}
where $S_{m_F}\equiv \sum_{m_F}\psi^{\dagger}_{m_F}({\bf 
r})h_{m_F}({\bf r})
\psi_{m_F}({\bf r})$, $S_{m_{F,F'}}\equiv 
\sum_{m_F,m_F'}\psi^{\dagger}_{m_F}({\bf r})
\psi^{\dagger}_{m_F'}({\bf r})\psi_{m_F'}({\bf r})\psi_{m_F}({\bf 
r})$, 
and 
$S_{m^{a,a'}_F,m^{b,b'}_F}\equiv\sum_{m^a_F 
m^{a'}_F}\sum_{m^b_F m^{b'}_F}\psi^{\dagger}_
{m^a_F}({\bf r})
\psi^{\dagger}_{m^{a'}_F}({\bf r})
{\bf F}_{m^a_F m^b_F}\cdot{\bf F}_{m^{a'}_F 
m^{b'}_F}\psi_{m^{b'}_F}({\bf r})
\psi_{m^b_F}({\bf r})$. 
In the equations above  
$h_{m_F}({\bf r})=-\hbar^2\nabla^2/2m+V_T({\bf r})$ is the single-atom 
Hamiltonian,  and $V_T({\bf r})$ is the 
trapping potential created by an optical field. 
Without loss of generality, we restrict ourselves to the  
case of hyperfine atomic state $F=1$ with the corresponding sub-levels: 
$m_F=-1,0,1$. 

The nonlinear interaction between different 
spin components of the condensate is governed by  
the spin-independent and spin-dependent interaction 
strengths, $c_0=4\pi\hbar^2(2a_2+a_0)/{3m}$ and $c_2=4\pi\hbar^2(a_2-a_0)/{3m}$, respectively. Here $a_0$ and $a_2$ are the s-wave 
scattering lengths 
for two  colliding atoms with total angular momentum $F_{{\rm tot}}=0$ and 
$F_{{\rm tot}}=2$. The coefficient $c_{2}$ is calculated to be {\em positive} for
the polar spinor condensate (e.g., $^{23}Na$) and {\em negative} for a ferromagnetic 
condensate (e.g., $^{87}Rb$) (see, e.g., \cite{ho}). 
Whether the spinor condensate is ferromagnetic or antiferromagnetic arises
from the fundamental gauge symmetry of the system, which in turn is dependent on 
the sign of the coefficient $c_2$ ($c_0$ is always positive for condensates
with repulsive interactions, such as the $^{23}Na$ and $^{87}Rb$ considered here).

From Eq. (\ref{Hamiltonian}) we can derive three coupled field 
equations for the case under consideration:
  \end{multicols}
  \noindent
  \begin{picture}(245,8)
	\put(0,0){\line(1,0){245}}
	\put(245,0){\line(0,1){8}}
  \end{picture}
\begin{equation}
 \label{firstset}
\begin{array}{l} {\displaystyle
i\hbar\frac{\partial \psi^{}_1}{\partial t}={\mathcal{L}}\psi^{}_1+
c^{}_2(\psi^{\dagger}_{1}\psi^{}_{1}+\psi^{\dagger}_{0}\psi^{}_{0}-\psi^{\dagger}_{-1}
\psi^{}_{-1})\psi^{}_{1}+c^{}_2\psi^{\dagger}_{-1}\psi^2_{0},}\\[7pt]
{\displaystyle
i\hbar\frac{\partial \psi^{}_{-1}}{\partial t}={\mathcal{L}}\psi^{}_{-1}+
c^{}_2(\psi^{\dagger}_{-1}\psi^{}_{-1}+\psi^{\dagger}_{0}\psi^{}_{0}
-\psi^{\dagger}_{1}\psi^{}_{1})\psi^{}_{-1}+c^{}_2\psi^{\dagger}_{1}\psi^2_{0},}\\[7pt]
{\displaystyle
i\hbar\frac{\partial \psi^{}_0}{\partial t}={\mathcal{L}}\psi^{}_0+
c^{}_2(\psi^{\dagger}_{1}\psi^{}_{1}+\psi^{\dagger}_{-1}\psi^{}_{-1})\psi^{}_{0}+
2c^{}_2\psi^{\dagger}_{0}\psi_{1}\psi_{-1},}\\
\end{array}\end{equation}

\begin{multicols}{2}

where ${\mathcal{L}}\equiv\left[-\frac{\hbar^2}{2m}\nabla^{2}+V_T({\bf 
r})\right]+c_0(\psi^{\dagger}_{-1}\psi_{-1}+\psi^{\dagger}_{0}\psi_{0}
+\psi^{\dagger}_{1}\psi_{1})$.
Next, we introduce the following linear field transformation 
for the quantum-field components, and treat the components as the 
corresponding mean fields:
\begin{equation}
 \label{symmetry}
\begin{array}{l} {\displaystyle
\phi_+\equiv 
\frac{1}{\sqrt{2}}(\psi_1+\psi_{-1}),}\\[7pt]
{\displaystyle
\phi_-\equiv 
\frac{1}{\sqrt{2}}(\psi_1-\psi_{-1}),}\\[7pt]
{\displaystyle
\phi_0\equiv \psi_0.}\\
\end{array}\end{equation}
As will become clear below, the advantage of this basis is that 
the stationary solutions for 
all three spinor components can be described by a single equation.

The coupled dynamical equations in the new basis have 
the following form:
\begin{equation}
 \label{secondset}
\begin{array}{l} {\displaystyle
i\frac{\partial\phi_+}{\partial t}={\mathcal{L}}\phi^{}_{+}+
c^{}_2\left((|\phi^{}_{-}|^2+|\phi^{}_{0}|^2)\phi^{}_{+}+(\phi_{-}^2+
\phi_{0}^2)\phi^*_+\right),}\\[7pt]
{\displaystyle
i\frac{\partial\phi_-}{\partial t}={\mathcal{L}}\phi^{}_{-}+
c^{}_2\left((|\phi^{}_+|^2+|\phi^{}_{0}|^2)\phi^{}_-+(\phi_{+}^2-
\phi_{0}^2)\phi^*_-\right),}\\[7pt]
{\displaystyle
i\frac{\partial\phi_0}{\partial t}={\mathcal{L}}\phi^{}_{0}+
c^{}_2\left((|\phi^{}_+|^2+|\phi^{}_{-}|^2)\phi^{}_0+(\phi_{+}^2-
\phi_{-}^2)\phi^*_0\right),}\\
\end{array}\end{equation}
where this time ${\mathcal{L}}\equiv\left[-\frac{1}{2}\nabla^{2}+V_T({\bf 
r})\right]+c_0(|\phi_+|^2+|\phi_0|^2+|\phi_{-}|^2)$.  The wave 
functions, time, spatial coordinates, and interaction strengths are 
measured in the units 
of $(\hbar/m\omega)^{-3/4}$, $\omega^{-1}$, 
$(\hbar/m\omega)^{1/2}$, and $(\hbar 
\omega)^{-1}(\hbar/m\omega)^{-3/2}$, respectively, 
where $\omega$ is the 
axial trapping frequency. 

{\em Stationary states.}The dynamics of the spinor condensate described
by  Eqs. (\ref{secondset}) is, in general, spin mixing.  However, any stable 
stationary solution of the system 
(\ref{secondset}) represents a non-spin-mixing, or 
spin-polarized state of the 
system. Such stationary states have a constant population of each spin component and
can be found by introducing the 
following ansatz: 
\begin{equation}\label{ansatz}
\phi_j=\sqrt{n_j({\bf r})}e^{-i\mu_{j}t+i\theta_j},
\end{equation}
where $j=(+,-,0)$, $\theta_{\pm}$ are the relative phases of each component with 
respect to $\phi_0(\theta_0\equiv0)$, and $\mu_{j}$ are the respective chemical 
potentials.  Finally, $n_++n_0+n_-\equiv n $ is the total density of 
the spinor condensate. 

Upon substitution of the ansatz (\ref{ansatz}) 
into 
Eqs. (\ref{secondset}), it becomes apparent that the stationary 
solutions 
can exist in this dynamical system only under the condition 
$\mu_{+}=\mu_{-}=\mu_{0}\equiv \mu$. Another condition is that the 
spin 
components of the condensate are {\em locked in phase}. 
For a stationary state to exist, the variables $2\theta_{-}$ and 
$2\theta_{+}$ can 
take only two distinct values, $0$ or $\pi$. In such a 
stationary state, the spinor eigenfunctions $\sqrt{n_j}$ are
 eigenmodes of the same effective potential created both by the optical 
trap and by the nonlinear interaction of all spin components.  Moreover, 
all three eigenmodes correspond to the 
same eigenvalue, $\mu$. This means that these eigenfunctions 
are 
proportional to each other and therefore can be presented in the 
form: $n_j({\bf r})=r_{j}n({\bf r})$, where constant coefficients 
$r_{j}$ represent the population of each spinor component in a steady 
state, with $r_{+}+r_{-}+r_{0}=1$. The spatial profiles of all three 
stationary spinor components 
obey the same time-independent equation:
\begin{equation}\label{stationary}
\left(-\frac{1}{2}\nabla^{2} -\mu+V_{T}({\bf r})\right)\sqrt{n({\bf r})}+\chi n({\bf r})\sqrt{n({\bf r})}=0.    
\end{equation}

Depending on the relative phases, there exist {\em four different 
phase-locked steady state solutions} of Eqs. (\ref{secondset}) with different populations in 
each 
spin component and different values of the coefficient $\chi$ in Eq. 
(\ref{stationary}):

{\em Case 1.} $e^{i\theta_{-}}=1$, $e^{i\theta_{+}}=i$.
 $r_{0}$ and $r_{\pm}$ are constrained by $r_{+}+r_{-}+r_{0}=1$, 
and $\chi=c_0$.

{\em Case 2.} $e^{i\theta_{-}}=1$, $e^{i\theta_{+}}=1$;
$r_{+}=r_{0}+r_{-}=1/2$, and $\chi=c_0+c_2$.

{\em Case 3.} $e^{i\theta_{-}}=i$, $e^{i\theta_{+}}=i$;
$r_{-}=r_{0}+r_{+}=1/2$, and $\chi=c_0+c_2$.

 {\em Case 4.} $e^{i\theta_{-}}=i$, $e^{i\theta_{+}}=1$;
$r_{0}=r_{-}+r_{+}=1/2$, and $\chi=c_0+c_2$.

{\em Modulational stability analysis.} 
The stationary solutions described above may correspond to different 
metastable states of the spinor condensate, provided that they are 
linearly stable. Linear instability, i.e. exponential growth of the 
modulation of the stationary {\em homogeneous} 
condensate, was previously hinted to be responsible for complex 
spatial modulations of the condensate in a trap, that ultimately lead 
to the destruction of 
the non-spin-mixing state \cite{pu}. However, no stability analysis 
of the spinor BEC has been carried out previously. Similar phenomenon occurring due 
to the nonlinear interaction of light beams (and pulses) in optical 
media is 
called {\em modulational instability} (MI) and is well studied in the 
context of nonlinear optical fibers \cite{mi}.

To perform the MI analysis for the spinor BEC,
 we first note that 
the stationary homogeneous (or constant density) solutions corresponding to the phase relations in the 
cases 1-4 
above have the form: $n_{j}^h=\mu/\chi$. Next, we add a small 
(generally 
complex) perturbation to the homogeneous solutions, taking the 
functions $\phi_{0}$ and $\phi_{\pm}$ in the form:
\begin{equation}\label{ansatz2}
\phi_j({\bf r},t)=(\sqrt{n_j^h}+\delta\phi_j)e^{i\theta_j-i\mu t}
\end{equation}
Substituting Eqs. (\ref{ansatz2}) into Eqs. (\ref{secondset}), 
omitting the terms containing $V_{T}$, and linearizing around the 
homogeneous solutions, we obtain the dynamical equations for the 
perturbations 
$\delta \phi_{j}({\bf r},t)$, which 
can be used to analyze the stability of the homogeneous solutions in 
the cases 1-4 against the growth of periodic perturbations. 
If the perturbations are taken in the form 
$\delta\phi_{j}=(u_{j}+iv_{j}){\rm cos}({\bf k}{\bf r})e^{\omega t}$, 
where ${\bf k}=(k_x,k_y,k_z)$,
these 
equations become: $\hat{A} \Omega^{T}=0$, where 
$\Omega=(u_{+},\,\,\,v_{+},\,\,\,u_{0},\,\,\,v_{0},\,\,\,u_{-},\,\,\,v_{-})$, 
the matrix $\hat{A}$ being too cumbersome to write out here. 
If the characteristic equation, $\det \hat{A}=0$, resolved with respect 
to the 
perturbation frequency $\omega$, has a real or complex root for some real 
positive ${\bf k}^{2}$, 
 the spinor condensate is modulationally unstable. In 
general, such an equation is of the sixth order in $\omega$. 
For simplicity, we can assume one of the populations $r_{j}$ 
equal 
to zero, and the other two equal to each other. Then for case 1, all 
possible eigenvalues are given by:
\begin{equation}\begin{array}{l}{\displaystyle
    \omega^{2}_{1}=-\frac{{\bf k}^{2}}{2}\left(\frac{{\bf k}^{2}}{2}+\mu 
\frac{c_{2}}{c_{0}}\right),}\\[7pt]{\displaystyle
    \omega^{2}_{2,3}=-\frac{{\bf k}^{4}}{4}-\mu 
\frac{{\bf k}^{2}}{2}\left(1+\frac{c_{2}}{c_{0}}\pm\sqrt{1
    +\frac{c_{2}}{c_{0}}}\right).}
    \end{array}
\end{equation}    
From these expressions, and keeping in mind that $|c_2/c_{0}|<1$, 
one can see that the real positive 
values of $\omega^2$, and hence the
MI of this solution, can occur {\em only for the ferromagnetic 
state}, i.e. when $c_{2}/c_{0}<0$.
 \begin{figure}
\setlength{\epsfxsize}{8.6cm}
\centerline{\epsfbox{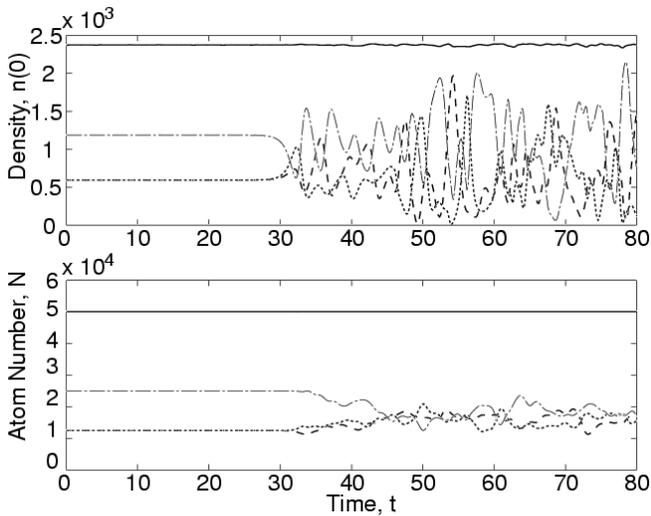}}
\caption[]{Time evolution of the central density $n(0)$ and atom number, N, of the
components of the 1D spinor condensate $\phi_-$ (dots) $\phi_{+}$ 
(dashed), $\phi_{0}$ (dot-dashed) and the total density and atom number (solid).
Parameters are $e^{i\theta_{-}}=1$, $e^{i\theta_{+}}=i$,
$r_{0}=1/2$, $r_{\pm}=1/4$, $\chi_0=0.0528$, $\chi_2=-0.00048$, and $\alpha=1$;
 $\phi_-|_{t=0}=\phi_{+}|_{t=0}$. The total number of atoms in the system is $N=5\times10^{4}$ corresponding to the 1D chemical potential $\mu\approx 125$.}
\label{fig1}
\end{figure}
Carrying out an identical analysis for 
 the cases 2-4, we find  
the following possible eigenvalues:
\begin{equation}\begin{array}{l}{\displaystyle
    \omega^{2}_{1}=-\frac{{\bf k}^{2}}{2},}\\[7pt]{\displaystyle 
 \omega^{2}_{2}=-2\left( \frac{{\bf k}^{2}}{2}-2\mu\frac{c_{2}}{c_0+c_{2}}\right)^{2},} \\[7pt]{\displaystyle
 \omega^{2}_{3}=-2\left(\frac{{\bf k}^{2}}{2}+2\mu\right).}
    \end{array}
\end{equation} 
Here all $\omega^{2}_j$ are negative and 
therefore, {\em MI does not occur} in any of 
the cases 2-4, neither for a ferromagnetic nor for a polar state.

It is possible to show that the conclusions of the MI analysis above hold
in the most general case, when none of the populations $r_j$ are zero. 
Thus {\em homogeneous solutions in the polar state never experience MI}.
\begin{figure}
\setlength{\epsfxsize}{8.6cm}
\centerline{\epsfbox{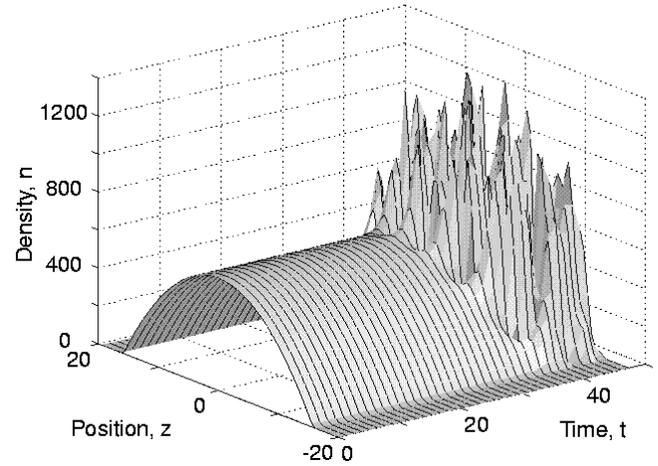}}
\caption[]{
Time evolution of the 1D $\phi_-(z)$ component for the case 1, showing the spatial development
 of the MI.  The $\phi_{+,0}(z)$ components behave similarly.
Parameters are as in Fig. \ref{fig1}.}
\label{fig2}
\end{figure}
{\em Numerical simulations.}
The MI analysis for the homogeneous condensate does 
not, strictly speaking, apply to the 
trapped condensate. However, it can serve as an indication of the 
spinor condensate behavior since the instability of the homogeneous 
condensate is bound to trigger the formation of the complex patterns in the trapped 
condensate cloud if the characteristic spatial extend of the condensate, 
$l$ (in dimensionless units), is larger than the largest length scale
 of the spatial patterns due to the MI, $L=k^{-1}_{{\rm min}}$. For the only case when the MI
of the spinor system does 
occur, this condition becomes $l > (\mu 
|c_{2}/c_{0}|)^{-1/2}$.

To test the results of our stability analysis and to demonstrate the 
effect of the MI on the trapped spinor condensate, we carry out 
numerical simulations of the dynamical equations (\ref{secondset}). 
Because of the anisotropic nature of the optical trapping potential, 
the cigar shaped BEC is assumed to be quasi-one-dimensional and hence we use the ansatz 
$\phi_j({\bf r},t)=\Psi(x,y)\phi_{j}(z,t)$, where $z$ is the direction 
of weak confinement, and $\Psi(x,y)$ is the 
 the wavefunction of the two-dimensional harmonic oscillator.  This leads us to
 the one-dimensional (1D) dynamical spinor 
system for $\phi_{j}(z,t)$ which is identical to the system (\ref{secondset}) with 
${\mathcal{L}}=\frac{1}{2}\left(-\partial^2 / \partial 
z^2+z^2\right)+\chi_0(|\phi_+|^2+|\phi_0|^2+|\phi_{-}|^2)$. 
The dimensionless interaction coefficients are 
$\chi_{0,2}=c_{0,2}\alpha$, 
and $\alpha=\int 
|\Psi(x,y)|^4dxdy/\int |\Psi(x,y)|^2dxdy$
is the transverse structure factor.
Using the initial 
conditions specified by: 
\begin{equation}\label{init_{cond}}
\phi_j(z)=\sqrt{r_{j}n(z)}e^{i\theta_j},
\end{equation}
 where $n(z)$ is the 1D spatial 
profile determined from  Eq. (\ref{stationary}),
we solve the 1D equivalent of Eqs. (\ref{secondset}) numerically for 
the cases 1-4 of the stationary phase-locked solutions. 
In agreement with the analytical results, we only observe MI for the ferromagnetic state ($c_2/c_0<0$) in the case 1.  The results of a 
representative calculation for the ferromagnetic state (corresponding to $^{87}Rb$)
 are shown 
in Figs. \ref{fig1} and \ref{fig2}. In all other cases the spinor BEC remains stable
to periodic modulation of its components.

\begin{figure}
\setlength{\epsfxsize}{8.6cm}
\centerline{\epsfbox{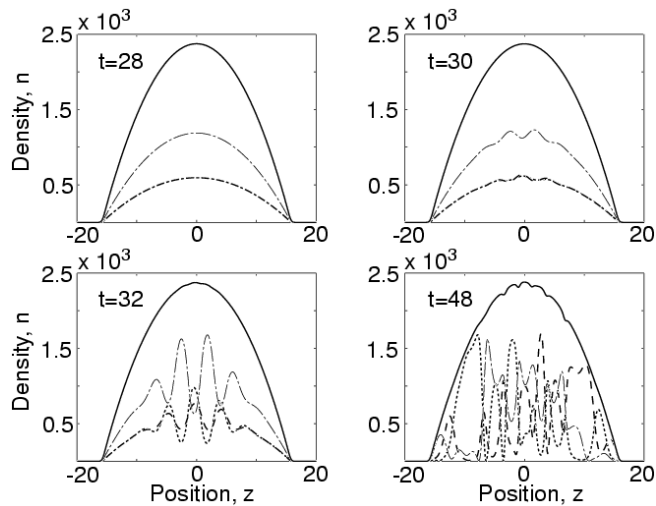}}
\caption[]{Spatial intensity profiles of the original spinor components, $\psi_j$,
demonstrating spin domain fragmentation. $\psi_-(z)$ (dots), $\psi_{+}(z)$
(dashed), $\psi_{0}(z)$ (dot-dashed) and the total density (solid).
Parameters are as in Fig. \ref{fig1}.}
\label{fig3}
\end{figure}

\begin{figure}
\setlength{\epsfxsize}{8.6cm}
\centerline{\epsfbox{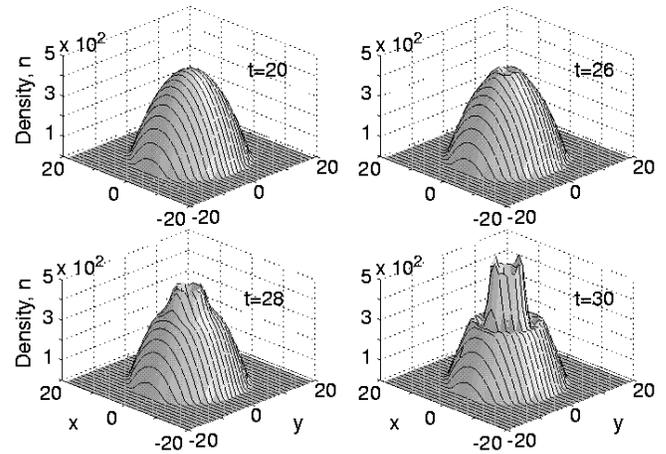}}
\caption[]{Developement of MI in the 2D case, shown for the
spatial intensity profile of the $\phi_+$ spin component.  Intial conditions are calculated from
Eq. (\ref{stationary}).  Parameters are as Fig. \ref{fig1}
 with the structure factor $\alpha_{2D}=1$. }
\label{fig4}
\end{figure}

Our results show that the effect of the MI on the trapped spinor BEC is 
twofold.  First, the instability destroys the non-spin-mixing stationary state, leading to population transfer between the spin components, as shown in Fig. (\ref{fig1}).  Second, the MI causes periodic spatial modulation of the condensate which, being confined by the trap, grows to be chaotic with time (see Fig. \ref{fig2}).
In the short term, MI causes the spatial fragmentation of the spin domains
 shown in Fig. \ref{fig3} for the original spinor components, $ \psi_j$.
Finally, to check that our results are dimensionally independent we 
have also numerically analyzed the spinor system in two dimensions, and again found that the MI 
only occurs in the ferromagnetic state for the stationary solution of case 1, as shown in Fig. \ref{fig4}.

In conclusion, we have predicted analytically and demonstrated 
numerically
 the possibility of MI of the ferromagnetic 
ground state
 of the spinor BEC.  This effect resembles the parametric MI in birefringent optical fibers with the Kerr 
nonlinearity, 
and it provides one more direct analogy between 
matter wave,
 and nonlinear, optics.  Our results show that the ferromagnetic spinor BEC is an 
ideal 
candidate for the first experimental observation of MI, spin domain fragmentation, and spatio-temporal chaotic dynamics
 of matter waves.

The authors gratefully acknowledge a reading of this manuscript by Dr. H. Pu.
This work has been partially supported by the Performance and Planning Fund of the Institute of Advanced Studies.

\end{multicols}
\end{document}